\documentclass[twocolumn,showpacs,preprintnumbers,aps,pre,showkeys]{revtex4}
\usepackage[T1]{fontenc}
\usepackage[latin1]{inputenc}
\usepackage{graphics}

\makeatletter

\providecommand{\LyX}{L\kern-.1667em\lower.25em\hbox{Y}\kern-.125emX\@}

\usepackage[T1]{fontenc}
\usepackage[latin1]{inputenc}
\usepackage{graphics}

\makeatletter

\usepackage[T1]{fontenc}
\usepackage[latin1]{inputenc}
\usepackage{graphics}

\makeatletter

\usepackage{natbib}
\usepackage{graphicx}
\usepackage{bm}
\usepackage{amsfonts}
\usepackage{amssymb}
\usepackage{amsmath}

\def\pd{\! \cdot \!}
\def\ta{{\bm \tau}}
\def\ep{{\bm \varepsilon}}

\def\iks{{\bm x}}
\def\igr{{\bm y}}
\def\ka{{\bm k}}
\def\eu{{\bm e}}

\makeatother
\makeatother
\makeatother
\makeatother
\makeatother
\makeatother

\makeatother
\makeatother

\makeatother
\makeatother
\makeatother
\makeatother

\begin{document}

\title{Fracture of disordered solids in compression as a critical phenomenon:\\ 
III. Analysis of the localization transition}

\author{Renaud Toussaint} \email[email: ]{Renaud.Toussaint@fys.uio.no} 
\altaffiliation[Present address: ]{ Department of Physics, University of Oslo, P.O. Box 1048 Blindern, 
0316 Oslo 3, Norway} 
\affiliation{G\'eosciences Rennes, Universit\'e de Rennes 1, 35042 Rennes Cedex, France}

\author{Steven R. Pride} \email[email: ]{Steve.Pride@univ-rennes1.fr} 
\affiliation{G\'eosciences Rennes, Universit\'e de Rennes 1, 35042 Rennes Cedex, France}

\date{\today}

\begin{abstract}

The properties of the Hamiltonian developed in Paper II are  studied showing  
 that at a particular strain level a  ``localization'' phase  transition occurs    
 characterized 
by the emergence of conjugate bands  of coherently 
oriented cracks. The functional integration  that yields 
the  partition function is then  performed 
 analytically using an approximation that employs only a subset of states in the 
functional neighborhood surrounding the most probable states. 
Such integration establishes the free energy of the system, and upon taking the derivatives of the 
free energy,  the localization transition is 
shown to be continuous and to be  distinct from  peak stress. 
When the bulk modulus of the  grain material is large,  
localization  always occurs in the softening regime  
following  peak stress, while  
for sufficiently small bulk moduli and at sufficiently low confining pressure, 
the localization  occurs in the hardening regime prior to  peak stress.
 In the approach to  localization,  the stress-strain relation
 for the whole rock remains analytic, as  is observed both in experimental data and in simpler models. 
  The correlation function of the crack fields is also obtained. It has  a correlation 
length  characterizing the aspect ratio of the crack clusters that  diverges    
as \( \xi \sim (\ep _{c}-\ep )^{-2} \) at  localization.

\end{abstract}


\maketitle

\section{Introduction\label{sec:intro} }

In Paper II of this series, we 
 obtained the
 Hamiltonian \( E_{j}(\ep ,\ep_{m}) \) of a population of interacting cracks 
which is the energy necessary to lead a mesovolume
of a disordered-solid  system from uncracked and unstrained initial conditions, to a final crack state
\( j \) at a maximum   imposed  strain \( \ep_{m} \) that is possibly different 
than the actual strain $\ep$ if the system has been subsequently unloaded.  
 Using this Hamiltonian, 
 we prove here that at a well-defined strain $\ep_c$,
the system undergoes a phase transition to bands of coherently oriented cracks.

To study the nature of this localization transition, we must evaluate the partition
function \( Z \) from which all physical properties depending on the crack
distribution are  obtained through differentiation. In Paper I, it was established
that \( Z \) takes a standard form 
\begin{equation}
\label{eq:part,function,principle}
Z(\ep ,\ep_{m},T)=\! \! \sum _{j}\! e^{-E_{j}(\ep ,\ep_{m})/T}
\end{equation}
 despite the fact that it derives from the initial quenched disorder in the
grain-contact strengths and has nothing to do with fluctuations through time.
 The possible crack states
\( j \) for a mesovolume are defined by a  local order parameter \( \varphi (\iks ) \)
 distributed at each cell \( \iks  \) of a regular square network of identical
cells. The amplitude of \( \varphi (\iks ) \) corresponds to the length of
a local crack (always less than cell dimensions), and its sign indicates
its orientation (\( \pm 45^{\circ } \) relative to the principle-stress axis).  

Our approach for performing the sum over  states begins   
by determining
which fields \( \varphi  \)  maximize the Hamiltonian.  Because 
the temperature in strain-controlled  experiments is negative, such 
 maximizing states are  
the dominant terms  in Eq.\ (1).
Any change in the nature of the maximizing crack-fields or in the nature of
the Hamiltonian in their neighborhood (\emph{e.g.}, the vanishing of a second derivative)  
 corresponds to  a phase transition. 

In Sec.\ II, the localization transition is identified and the geometrical nature  of 
the crack fields in the ``functional neighborhood'' surrounding  the 
maximizing states  defined.   In Sec.\ III,  we sum only over this 
subset of all states to obtain an  analytical approximation of \( Z \).
In Sec.\ IV, the free energy \( F=-T\ln Z \) is differentiated with respect 
to \( {\bm \ep } \) and  \( T \)  to determine both the sustained 
stress \( \ta  \),   the energy \( U \), and  the entropy $S$. 
In the approach to localization, no singularities are present in either $F$ or 
any of its derivatives with respect to strain or temperature which demonstrates, among other things, 
 that the stress/strain 
relation is analytic up to (and including) 
localization.  In  Sec.\ V,  an external field $J$ is introduced 
that couples to $\varphi$  permitting an autocorrelation function to be obtained. 
All  singularities  at localization  are in  the second (and higher) 
 derivatives of \( F \) with respect to $J$ with the consequence that   
the  correlation length  diverges  as 
 \( \xi \sim (\ep _{c}-\ep )^{-2} \). 

\section{Principle of the transition\label{sect:principle,of,transition}}
\subsection{Extrema of the Hamiltonian\label{max,hamilt}}

We  now determine the most probable states by maximizing the  
Hamiltonian \( E_{j}(\ep ,\ep_{m}) \) along the load path  
 $\ep = \ep_m$.  From the summary of Paper II, we have   
\[
E_{j}=E^{0}(\ep_{m})+(1-q)\left\{E^{\text {av}}(\ep_{m})[\varphi ]+E^{\text {int}}(\ep_{m})[\varphi ] 
\right\}\]
where $E^0$ is the energy of the intact material, $E^{\rm av}$ is the energy due 
to the crack field when crack interactions are neglected, and $E^{\rm int}$ is 
the energy due to crack interactions.  The parameter $q$ derives from the quenched-disorder 
distribution and is bounded as $1/2 \le q < 1$.

That the  Hamiltonian must be maximized and not minimized comes from
the  temperature parameter being negative as  was quantitatively established in 
Sec.\ IV of Paper I.  
Because we assume the system is intact before strain is applied, it is a fact 
of our model that the intact state is always the most probable.  
For this to hold true, the temperature must be negative in strain-controlled 
experiments because the arrival of cracks at constant strain always reduces 
the  energy in a mesovolume. 

\subsubsection{Mean-field terms\label{eq:mean,field,term}}

A mean-field simplification of the model built in Paper II would reduce the
Hamiltonian to the sole term 
\begin{eqnarray*}
E^{0}+(1-q)E^{\textrm{av}}&=& \frac{1}{2}\left[ \alpha \Delta^2
 +(1-\alpha ) \gamma^2\right] \\
&& -(1-q)\epsilon  \overline{\psi} \left[\kappa_2 \Delta^2  +
 \kappa_3 \gamma^{2}\right]
\end{eqnarray*}
where  $\Delta$ is the strain dilatation, $\gamma$ the  
shear strain, 
and   $\alpha$ and $\kappa_i$ are combinations of the elastic moduli all as defined in Paper II.
The  second term is strictly negative and represents  
the weakening of the rock due to the  crack porosity  which is proportional to  \( \overline{\psi } \),  
 the volume average of  
 the positive  field \( \psi =\left| \varphi \right|\).
Therefore this mean-field
Hamiltonian is maximum when \( \overline{\psi }=0 \), which uniquely 
corresponds to
 the uniform intact state \( \psi =\varphi =0 \).

\subsubsection{Interaction term\label{sect:long,range,elast}}

The interesting term is the interaction energy 
\( E^{\textrm{int}} \).   As defined in the summary of Paper II (the reader should consult 
this summary for the definition of all the terms in what follows), $E^{\rm int}$  
 is a sum
over wavenumbers \( \ka  \) of orthogonal quadratic forms involving \( R_{\ka } \)
and \( I_{\ka } \), which are vectors containing  the \(\ka \)-space Fourier modes of the
order-parameter fields \( \varphi  \) and \( \psi  \). 
The sign of these  forms is determined by  
 the sign of the two eigenvalues of the symmetric matrices \( P_{\ka } \).
For any \( \ka  \), at least one of the eigenvalues
is positive, since \( [1,0]\cdot P_{\ka }\cdot [1,0]^{T} 
=L_{\ka }=\Delta ^{2}\kappa ^{2}_{1}(1-\alpha u^{2}_{\ka })>0 \)
where \(1/2 <  \alpha < 1  \)  and \( u_{\ka } \) is a cosine.
To determine the sign of the second eigenvalue, it is sufficient to take the
determinant of \( P_{\ka } \). Using \( u^{2}_{\ka }+v^{2}_{\ka }=1 \), it
is straightforward to show that 
\begin{equation}
\label{eq:det.P1}
\det |P_{\ka }|=\Delta ^{4}\kappa ^{4}_{1}(1-\alpha )\left[ c  
v_{\ka }+\omega \right]^{2}. 
\end{equation}
 This is strictly positive for every \( \ka  \), except when 
\begin{equation}
\label{eq:cond,k}
v_{\ka }=\sin (2\theta _{\ka })=-  \omega /c 
\end{equation}
 in which case the determinant and second eigenvalue are zero. The vanishing
of the determinant is thus independent of the norm of \( \ka  \), and takes
place at either of two conjugate angles \( \theta _{\ka }^{+}=\arcsin (-\omega /c)/2 \)
or \( \theta _{\ka }^{-}=\pi /2-\arcsin (- \omega /c )/2 \) where \( \theta _{\ka } \)
represents the angle between \( {\ka } \) and the crack-orientation vector
\( \hat{\eu }_{1} \). The directions in \( {\ka } \)-space at which the determinant
vanishes will be denoted by the unit vectors \( \hat{\ka }^{\pm } \). Thus,
the matrices \( P_{\ka } \) are positive definite; \emph{i.e.}, they have two
strictly positive eigenvalues, except for those particular wavevectors lying
along one of the two directions for which they become positive degenerate. The
eigenvector of \( P_{\ka } \) associated with the zero eigenvalues is easily  
computed to be \( [1, \, -M_{\ka }/L_{\ka }]^{T} \).  

The positive-definite  quadratic
forms of $E^{\rm int}$  are multiplied by a negative constant which  implies
that the maximum of \( E^{\textrm{int}} \) occurs when \( \widetilde{\varphi }_{\ka }=\widetilde{\psi }_{\ka }=0 \)
for every non-zero \( \ka  \) with the exception of those \( \ka  \) satisfying
Eq.\ (\ref{eq:cond,k}). At these degenerate angles, the Fourier modes of \( \varphi  \)
and \( \psi  \) are related as 
\begin{equation}
\label{eq:zero,eigenvect}
\widetilde{\varphi }_{\ka }=-\frac{L_{\ka }}{M_{\ka }}\widetilde{\psi }_{\ka }.
\end{equation}
 Now, the definition of the auxiliary field \( \psi _{x}=\left| \varphi _{x}\right|  \)
imposes a series of constraints between \( \widetilde{\varphi }_{\ka } \) and
\( \widetilde{\psi }_{\ka } \). The simplest is obtained by noting that the
space-integrals of \( \varphi ^{2} \) and \( \psi ^{2} \) must be the same  which
is equivalent to 
\begin{equation}
\label{eq:rel,Fourier,phi,psi}
\sum _{\ka }\left( \widetilde{\psi }_{\ka }\widetilde{\psi }_{-\ka }
-\widetilde{\varphi }_{\ka }\widetilde{\varphi }_{-\ka }\right) =0.
\end{equation}
 For a crack-state maximizing \( E^{\textrm{int}} \), this condition further
requires that 
\begin{equation}
\label{eq:ctrainte,dir,part}
\left( \tilde{\varphi }^{2}_{0}-\tilde{\psi }^{2}_{0}\right) +\! \! \! \sum _{\begin{array}{c}
\ka =k\hat{\ka }^{\pm }\\
k\neq 0
\end{array}}\! \! \!\!
\left( 1-\frac{M_{\ka }^{2}}{L_{\ka }^{2}}\right) 
\widetilde{\varphi }_{\ka }\widetilde{\varphi }_{-\ka }=0. 
\end{equation}
 It will be seen momentarily that along the directions \( \hat{\ka }^{\pm } \),
the factors \( 1-M_{k\hat{\ka }^{\pm }}^{2}/L_{k\hat{\ka }^{\pm }}^{2} \) are
equal, and that this quantity is an increasing function of the shear-strain
parameter \( \omega = (\kappa_3 \gamma)/(\Delta \kappa_1) \), 
starting at a strictly negative value when \( \omega =0 \)
(no shear deformation yet applied), and reaching \( 0 \) at a particular value
\( \omega _{c} \). For every wavevector, \( \widetilde{\varphi }_{\ka }\widetilde{\varphi }_{-\ka }
=\left\Vert \widetilde{\varphi }_{\ka }\right\Vert ^{2} \)
is trivially positive, and the definition of \( \psi  \) also requires that
\( \tilde{\varphi }^{2}_{0}-\tilde{\psi }^{2}_{0}\leq 0 \) for any crack state.
 From Eq.\ (\ref{eq:ctrainte,dir,part}), we can conclude that for \( \omega <\omega _{c} \),
the only crack-states maximizing the interaction term \( E^{\textrm{int}} \)
must satisfy both \( \tilde{\varphi }^{2}_{0}=\tilde{\psi }^{2}_{0} \) and,
for every non-zero \( \ka  \), \( \widetilde{\varphi }_{\ka }=\widetilde{\psi }_{\ka }=0 \).
Such a maximum thus corresponds to a spatially uniform crack field. 

At the degenerate
point \( \omega =\omega _{c} \), the set of maximizing crack states goes through
a drastic change.   Any non-zero Fourier mode of \( \varphi  \) and \( \psi  \)
along the directions \( \hat{\ka }^{\pm } \) no longer modifies \( E^{\textrm{int}} \)
so long as \( \tilde{\varphi }^{2}_{0}=\tilde{\psi }^{2}_{0} \); \emph{i.e.},
so long as the crack field has  the same sign over the entire mesovolume. This
degeneracy of \( E^{\textrm{int}} \) at \( \omega =\omega _{c} \) is at the
origin of the localization  phase transition.

The critical value \( \omega _{c} \), and the corresponding wavevectors \( \ka  \)
for which non-zero Fourier modes of \( \varphi  \) and \( \psi  \) do not
contribute to \( E^{\textrm{int}} \), are determined from the two conditions
\begin{eqnarray}
\det (P_{\ka }) & = & 0\label{eq:det,P,zero} \\
L_{\ka }^{2}-M_{\ka }^{2} & = & 0.\label{eq:Lsquare,minus,Msquare,zero} 
\end{eqnarray}
 Using the solution of Eq.\ (\ref{eq:det,P,zero}) given by Eq.\ (\ref{eq:cond,k})
in the definitions of \( L_{\ka } \) and \( M_{\ka } \) given in the summary of Paper II, 
 Eq.\  (\ref{eq:Lsquare,minus,Msquare,zero}) then becomes an equation
for \( \omega _{c} \)
\begin{equation}
\label{eq:Lcarre,moins,Mcarre}
\left[ \omega _{c}^{2}-\left(c^2 - 1 \right)\right] 
\left[ \omega _{c}^{2}+\frac{(1-\alpha )}{\alpha }c^2 
 \right] =0.  
\end{equation}
 From the definitions of Paper II, 
we have  $c>1$ while $1/2< \alpha < 1$.  
 Thus, Eqs.\ (\ref{eq:det,P,zero})--(\ref{eq:Lsquare,minus,Msquare,zero})
can only be satisfied by 
\begin{eqnarray}
\omega =\omega _{c}^{\pm } & = & \pm \sqrt{c^2 -1}\label{eq:omegac} \\
\sin (2\theta _{\ka }) & = & -\left(\sqrt{c^2-1 }\right)/c.  \label{eq:kac} 
\end{eqnarray}
 With a radial confining pressure maintained constant, and a positive shear
stress \( \tau _{\text {axial}}>\tau _{\text {radial}} \), the strain components
of the rock satisfy  \( \varepsilon _{\text {axial}}<\varepsilon _{\text {radial}} \)
and \( \varepsilon _{\text {axial}}<0 \) so that  
\( \omega =(\kappa _{3}/\kappa_1) 
(\varepsilon _{\text {axial}}-\varepsilon _{\text {radial}})/
(\varepsilon _{\text {axial}}+\varepsilon _{\text {radial}}) \)
is a positive and monotonically increasing function of the axial stress, until
the rock possibly exhibits some positive volumetric strain (we will later show that this 
does not occur prior to localization), 
 where this quantity diverges to \( +\infty  \) and increases further
starting from \( -\infty  \).  All of this establishes that Eqs.\ (\ref{eq:det,P,zero})
and (\ref{eq:Lsquare,minus,Msquare,zero}) have no solution until the first
solution   \( \omega =\omega ^{+}_{c} \) is reached. At this particular
strain value, non-zero Fourier modes of \( \varphi  \) and \( \psi  \) having
any wavevector lying in one of the two directions defined by Eq.\ (\ref{eq:kac})
can be added to a mesovolume with no change in the interaction energy. 

For quartz as the rock mineral, 
$$ \frac{\kappa_1}{\kappa_3} \omega ^{+}_{c} 
=  \left(\frac{\varepsilon _{\text {axial}}-\varepsilon _{\text {radial}}}{
\varepsilon _{\text {axial}}+\varepsilon _{\text {radial}} }\right)_{\!\!c}  
\simeq 12,$$ 
so that we find   
 \( (\varepsilon _{\text {axial}}/\varepsilon _{\text {radial}})_{ c} \simeq -1.2 \) at the 
transition.  Our model thus predicts the localization transition   
 to occur after  a sign reversal
of \( \varepsilon _{\text {radial}} \) but     prior to the point   
where  \( \Delta =\varepsilon _{\text {axial}}+\varepsilon _{\text {radial}} \) changes sign.
These  results are consistent    
with what is observed in usual triaxial
mechanical experiments ({\em e.g.}\  \cite{BPS66,SHS73,Ui76}).

It can now be algebraically  verified using the definitions of $L_{\bm k}$ and $M_{\bm k}$ given 
in Paper II, that \( 1-M_{k\hat{\ka }^{\pm }}^{2}/L_{k\hat{\ka }^{\pm }}^{2} \)
does not depend on the norm \( k \) nor on which of the two directions \( \hat{\ka }^{\pm } \)
is selected. Further, it increases monotonically from a negative value to reach
zero when \( \omega =\omega_c^{+} \) (facts  used in obtaining the
above results).

\subsection{Structure at the localization transition\label{sect:presence,of,transition}}

The goal here is to define the geometric nature of the states 
maximizing \( E^{\textrm{int}} \) at the strain point \( \omega _{c} \).
Necessary conditions on the structure of the degenerate states were just given 
and these are easily made into sufficient conditions. First, the degenerate states
must correspond to crack fields of constant sign. 
They thus satisfy everywhere
\( \psi =\varphi  \) or \( \psi =-\varphi  \) or, 
equivalently, \( \widetilde{\psi }_{\ka }=\widetilde{\varphi }_{\ka } \)
or \( \widetilde{\psi }_{\ka }=-\widetilde{\varphi }_{\ka } \). Considering
this together with the necessary conditions of Eqs.\ (\ref{eq:zero,eigenvect})
and (\ref{eq:det,P,zero}), requires that the degenerate states be one of two
types: (1) \( \varphi >0 \) everywhere and the only possible non-zero Fourier
modes of \( \varphi  \) have wavevector directions that satisfy \( N_{\ka }/M_{\ka }=M_{\ka }/L_{\ka }=-1 \);
or (2) \( \varphi <0 \) everywhere and the wavevector directions satisfy \( N_{\ka }/M_{\ka }=M_{\ka }/L_{\ka }=+1 \).
Using again the definitions of \( L_{\ka },\, M_{\ka },\, N_{\ka } \) given
in the summary of Paper II, the first type of degenerate mode corresponds to
wavevectors satisfying \( \sin (2\theta _{\ka })=-\sqrt{c^2-1}/c \)
and \( \cos (2\theta _{\ka })=-1/c \), while the second
type of mode has the same sine requirement, but an opposite value for the cosine.
Using \( \hat{\ka }^{+} \) to represent the wavevector direction corresponding
to the first condition, and \( \hat{\ka }^{-} \) the wavevector direction for
the second condition, we conclude that the emergent degenerate crack states
consist either of right-inclined cracks with spatial fluctuations forming bands
perpendicular to \( \hat{\ka }^{+} \), or of left-inclined cracks forming bands
perpendicular to \( \hat{\ka }^{-} \). Such geometry is sketched in Fig.\  \ref{fig:orientation,bandes,conjuguees}. 
\begin{figure}
{\par\centering \resizebox*{6cm}{5.5cm}{\includegraphics{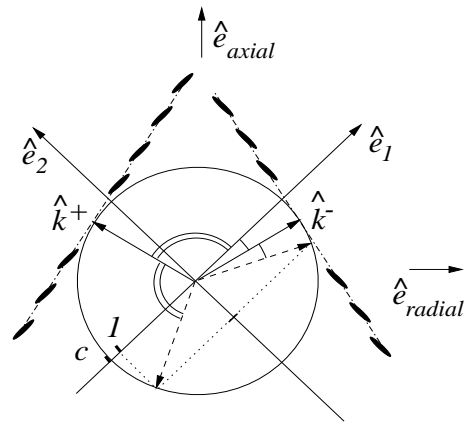}} \par}
\caption{A part of the conjugate bands emerging at the critical strain.
\label{fig:orientation,bandes,conjuguees} The
bands perpendicular to $ \hat{\bm k}^{+}$ 
are exclusively composed of right-inclined cracks, while those perpendicular to \
$\hat{\bm k}^{-}$ contain only left-inclined cracks. } 
\end{figure}

These two sets of crack modes are conjugate to each other; \emph{i.e.}, symmetric
to each other under inversion of the radial axis. Since they become statistically
important as \( \omega \rightarrow \omega _{c} \), whereas the intact state
or uniform states are the important states prior to \( \omega _{c} \), 
 the system spontaneously breaks its symmetry at the transition, which is
characteristic of a continuous phase transition.

Further, the angle formed by these bands is  at \( 45^{\circ }-\left| \theta _{\hat{\ka }^{-}}\right|  \)
from the axial direction. Using Eq.\ (\ref{eq:kac})
and the definitions of \( \kappa _{1} \) and \( \kappa _{2} \) in terms of
the Lam\'{e} parameters, it is found that 
 this angle is typically between \( 15^{\circ } \) and \( 35^{\circ } \)
depending on the rock mineral \cite{Bas95} considered which is consistent with laboratory experiments.

Finally, we note that these special crack bands that leave $E^{\rm int}$ unchanged,  
make a negative contribution to the Hamiltonian  through  the mean-field
energy \( {E^{\text {av}}} \) that is proportional to $\overline{\psi }$.
Due to the $r^{-D}$ range of elastic interactions, $E^{\rm int}$ is independent of 
the norm of ${\bm k}$ (it depends only on its orientation).  Thus, the spatial variation 
of the bands perpendicular to their lateral extent has no influence on $E^{\rm int}$; it 
only affects $E^{\rm av}$ through the number of cracks present.  
For large
systems and  a narrow band of only  a few cell widths, 
 \( \overline{\psi }=\Lambda \ell /\ell ^{2}=\Lambda /\ell  \)
where  \( \Lambda  \) and \( \ell  \) are the linear size of
a cell (grain) and of a mesovolume.   Thus, such a thin band 
makes a negligible  contribution to $\overline{\psi}$ 
   for large systems, and
is   energetically equivalent
to the intact state. However, states with numerous  and/or wide bands can make a non-negligible 
contribution to 
 \( \overline{\psi } \) and are, therefore, less probable.  
So this transition indeed corresponds to  ``localized''  structures.  
Only those   states with a small number of small width bands along the special  
directions are the statistically emergent ones as  is  observed in actual 
experiments on rocks.

\section{Obtaining the partition function}

\label{sect:approx,part,funct}

The sum over crack states in Eq.\ (\ref{eq:part,function,principle}) is equivalent
to the functional integration 
\begin{equation}
\label{eq:part,function,functional}
Z=\int \left( \prod _{\iks \in \Omega }d\varphi _{\iks }\right) e^{-E[\varphi ,\ep ,\ep_{m}]/T}.
\end{equation}
 Since our Hamiltonian is expressed in terms of the Fourier modes \( \tilde{\varphi }_{\ka } \),
it is shown in standard textbooks \cite{LB90b,Gol92} that \( Z \) further
transforms to 
\begin{equation}
\label{eq:Z,Fourier}
Z=\int d\tilde{\varphi }_{0}\underset {\ka \in \Upsilon }{\prod }(d\tilde{\varphi }^{R}_{\ka }d\tilde{\varphi }^{I}_{\ka })e^{-E[\tilde{\varphi }_{\ka },\ep ,\ep_{m}]/T}
\end{equation}
 where \( \widetilde{\varphi }^{R}_{\ka } \) and \( \widetilde{\varphi }^{I}_{\ka } \)
are the real and imaginary part of \( \widetilde{\varphi }_{\ka } \), and \( \Upsilon  \)
is a half-space of the set of the wavevectors corresponding to the non-zero
modes; \emph{i.e.}, corresponding in 2D to the discrete set \( (k_{1},k_{2})=({2\pi }{\ell }/n_{1},
{2\pi }{\ell }/n_{2}) \)
with \( (n_{1},n_{2})\in \mathbb {Z}^{2} \). There is a small-wavelength cutoff
given by \( \max (|n_{1}|;|n_{2}|)<\ell /\Lambda  \) that ensures that \( \varphi  \)
does not vary on scales smaller than that of a cell, and there is the arbitrary
criterion \( k_{1}>0 \) made to divide this space into two symmetrical parts.
Equation (\ref{eq:Z,Fourier}) is valid up to a multiplicative constant that
has no physical importance since the properties of a system correspond to the
derivatives of the free energy \( F=-T\ln Z \).

An analytic approximation for \( Z \) is obtained by performing the functional
integration over a properly chosen subset of all the possible crack-states.
The definition of this subset is based on what was learned in the preceeding
section; namely, that among the states having a given non-zero crack occupation
 \( \overline{\psi } \), the most probable are the uniform states, and
precisely at the phase transition,  certain banded states  may arrive
at almost no energy cost,  and  these emergent states  also have  
the same sign over space. Thus, the geometrical characteristic of all such
states in the ``functional neighborhood'' of the minimizing state is that
in each one, all cracks are oriented in the same direction (either left or right).
This property justifies making a so-called ``constant-sign'' (or ``mean-phase'')
approximation for the partition function in which only those states in which
the sign does not change in space will be considered. This still includes a
huge range of states in which \( \varphi  \) spatially varies. The excluded
states in this approximation are guaranteed to have lower probabilities than
the included ones and, as such, should have a negligible influence on the physical
properties of the system.
In this aproximation, the Fourier modes of the auxiliary \( \psi  \)
field are trivially related to those of \( \varphi  \) 
as either \( \widetilde{\psi }_{\ka }=\widetilde{\varphi }_{\ka } \)
for the positive states, or \( \widetilde{\psi }_{\ka }=-\widetilde{\varphi }_{\ka } \)
for the negative states.  

We now rescale the temperature as \( T=\Lambda^{D}T'/\ell ^{D} \).  
 From the definition $T= \partial U/\partial S$ and the fact that $U$ is an energy 
density independent  of $\ell$ while $S$ is extensive and thus increases as $\ell^D$, 
we have that $T$ scales as $\ell^{-D}$.  In taking the thermodynamic limit in what 
follows, it is convenient to work with the purely intensive parameter 
 \( T' \) (that is independent of \( \ell \) ). 
Our partition function within the constant-sign approximation then takes the form
\begin{eqnarray}
{Z\!\simeq \!\! \! \int _{+}\!\!\!\mathcal{D}\varphi \exp \! 
\left\{ \! \frac{-\ell ^{D}}{\Lambda ^{D}T'} \! 
\left[ d\!+\!e\!\left( \frac{\tilde{\varphi }_{0}}{\ell ^{D}}\right)\!\!  
+ \!\!\sum _{\ka \in \Upsilon }\!\! 
w^{+}\!(\ka )\!\left| \frac{\tilde{\varphi }_{k}}{\ell ^{D}}\right| ^{2}\right]\! \right\} } 
&& \nonumber \\
\!\!\!\!\!\!\!\!  +\! \! \int _{-}\! \! \! \mathcal{D}\varphi \exp \! \left\{ \! \frac{-\ell ^{D}}
{\Lambda ^{D}T'}\! \left[ d\!+\!e\!\left( \frac{\tilde{\varphi }_{0}}{\ell ^{D}}\right)\!\! 
+\! \! \sum _{\ka \in \Upsilon }\!\! w^{-}\!(\ka )\!
\left| \frac{\tilde{\varphi }_{k}}{\ell ^{D}}\right| ^{2}\right]\! \right\} &&
\label{eq:Z,negatifs}
\end{eqnarray}
where \( \mathcal{D}\varphi  \) is a compact notation for the functional measure  
\( d\tilde{\varphi }_{0}\prod_{k\in \Upsilon }(d\tilde{\varphi }^{R}_{k}d\tilde{\varphi }^{I}_{k}) \),
and where  \( \int _{+} \) and \( \int _{-} \) represent integration
over the subsets of \( \varphi  \) fields that are everywhere either positive or negative.
The quantities \( d,\, e \) and \( w^{\pm } \) are defined in the summary of Paper II as 
\begin{eqnarray}
d & = & \frac{1}{2}\left[ \alpha \Delta ^{2}+(1-\alpha )\gamma ^{2}\right] \\
e & = & -\frac{\epsilon }{2}\left[ \kappa _{2}(\Delta ^{2}-q {\Delta_m^{2}})+\kappa _{3}(\gamma ^{2}-
q {\gamma_m^{2}})\right] \\
w^{\pm }(\ka ) & = & -\frac{\epsilon ^{2}}{(1-\alpha )}
\left[ \left( L_{\ka }\pm 2M_{\ka }+N_{\ka }\right) (\ep )\right. \nonumber \\
 &  & \quad \qquad 
\left. - q \left( L_{\ka }\pm 2M_{\ka }+N_{\ka }\right) 
(\ep_m)\right]. \label{w+definition}
\end{eqnarray}
Recall that the values of the actual strain \( \ep  \)  intervening in
the probability distribution and in the partition function are those along the
load curve for which $\ep = \ep_m$. Their formal distinction only plays a role when  
partial derivatives of the free energy are taken to define stress. 
We note then that the value  
of \( w^{\pm } \) at \( \ep =\ep_m \) is \( w^{\pm }=-(1-q)\epsilon ^{2}[1,\pm 1]\! \cdot \! P({\ka })\! 
\cdot \! [1,\pm 1]^T/(1-\alpha ) \)
and since we have shown that \( P \) is a positive-definite matrix, and that
the  temperature \( T' \) is negative, we have that \( w^{\pm }/T' \)
 in Eq.\ (\ref{eq:Z,negatifs}) is strictly positive. 
 
The symmetry
of the problem under the parity transformation (inversion of the radial axis) 
guarantees that both integrals in
Eq.\ (\ref{eq:Z,negatifs}) are equal. Accordingly, only the first integral
over positive crack states will be treated. 
This  integral   
separates  into products of Gaussian  integrals with 
 the only remaining coupling between the Fourier modes
coming  from the complicated constraints on the integration domain boundaries
that are what guarantee \( \varphi  \) to have the same sign everywhere in
real space, and  \( \psi \) to lie within \( [0,1] \). But in order to
study any singular behavior of the free energy \( F \) in the vicinity
of localization, \( Z \) is determined in the thermodynamic
limit in which both the system size and  mesovolume size \( \ell  \)
are taken to be infinite. In this limit, the complicated integration bounds
in \( \ka  \)-space are not relevant. The integration can be carried out entirely
on \( \mathbb {R}^{+} \) for \( \widetilde{\varphi }_{0}/\ell ^{D} \), and
\( \mathbb {R} \) for each of the variables \( \widetilde{\varphi }^{R}_{\ka }/\ell ^{D} \),
\( \widetilde{\varphi }^{I}_{\ka }/\ell ^{D} \) without changing the result because 
the contribution to these integrals in the thermodynamic limit comes from  
 the immediate neighborhood of \( \widetilde{\varphi }_{0}/\ell ^{D}=0 \)  
and \( \widetilde{\varphi }_{\ka }/\ell ^{D}=0 \).  

A  technical proof of
this can be obtained as   follows: using \( \mathbb {R}^{+} \) and
\( \mathbb {R} \) as the integration domains  produces an upper bound
for \( Z \) since this includes every positive crack-field. A lower bound can
be obtained by reducing the integration domain to a subset of the set of all
positive crack-fields in which \( 0<\widetilde{\varphi }_{0}<\ell ^{D} \) and
\( \sum _{\ka \in \Upsilon }\left( \left| 
\tilde{\varphi }^{R}_{\ka }\right| +\left| \tilde{\varphi }^{I}_{\ka }\right| \right) 
\leq \min [\tilde{\varphi }_{0}/\sqrt{2},(\ell ^{D}-\tilde{\varphi }_{0})/\sqrt{2}] \).
Integrating mode by mode over this polyhedra, the result can be shown to be
asymptotically equivalent to the result of the upper bound in the limit where
\( \ell ^{D} \) becomes infinite. This  exercise is left to
the attention of the reader.

Thus, no coupling between the \( \ka  \) modes exists in the thermodynamic
limit, and our approximation of the partition function takes the convenient
form 
\begin{equation}
\label{eq:Z,reduced}
Z\simeq 2z_{0}z_{1}\underset {\ka \in \Upsilon }{\prod }\left[ z(\ka )^{2}\right] 
\end{equation}
 where 
$z_{0}  =  e^{-\ell ^{D}d/\Lambda ^{D}T'}$ and 
\begin{eqnarray}
z_{1} & = & \int _{x\in \mathbb {R}^{+}}dx\, e^{-{\ell ^{D}}ex/\Lambda ^{D}T'}\label{eq:z0} \\
z(\ka ) & = & \int _{x\in \mathbb {R}}dx\, e^{-{\ell ^{D}}w^{+}(\ka )x^{2}/\Lambda ^{D}T'}.\label{eq:zka} 
\end{eqnarray}
 In the limit \( \ell \rightarrow +\infty  \), these two integrals become  
\begin{eqnarray}
z_{1} & \sim  & \Lambda ^{D}T'/\left(\ell ^{D}e\right) \label{eq:z0.asympt} \\
z(\ka ) & \sim  & \sqrt{\pi \Lambda ^{D}T'/\left[\ell ^{D}w^{+}(\ka )\right]}.\label{eq:zka,asympt} 
\end{eqnarray}
Using Eq.\ (\ref{eq:Z,reduced}), one then obtains the free-energy density 
 in the thermodynamic limit  
\begin{eqnarray}
F & = & -T'(\ln Z)\Lambda ^{D}/\ell ^{D}\nonumber \\
 & \sim  & d+\frac{\Lambda ^{D}T'}{\ell ^{D}}\sum _{\ka \in \Upsilon }\ln 
\left( \frac{\ell ^{D}w^{+}(\ka )}{\Lambda ^{D}T'}\right). \label{eq:F,thermod,lim} 
\end{eqnarray}
 The contribution \( z_{1} \) has vanished in this limit due to the fact that
\( x\ln x\rightarrow 0 \) as \( x\rightarrow 0 \). This is a technical consequence
of the fact that for states composed of a few single bands, \( \bar{\psi } \)
vanishes in the thermodynamic limit, as  commented upon in  the previous section.


\section{System properties at localization}
The remaining task is to link this free energy to the observables
of the system by taking the partial derivatives of $F$ 
 in the limit as localization is approached.

The two partial
derivatives of primary interest are those that give the dimensionless
entropy density
$s=\Lambda^D S/\ell^D$ and the stress ${\bm \tau}$.  From Paper I, we have  
$$
-s= \left.\frac{\partial F}{\partial T'} \right|_{{\bm \varepsilon},
{\bm \varepsilon}_m}  \quad \mbox{ and } \quad
{\bm \tau} = \left.\frac{\partial F}{\partial {\varepsilon}} \right|_{T',
{\bm \varepsilon}_m}.
$$
The free energy  of Eq.\ (\ref{eq:F,thermod,lim}) is rewritten by replacing
the sum over
the wavevectors \( \sum _{\ka \in \Upsilon } \) with  a continuous
integral \( \ell ^{D}/(2\pi )^{D}\int _{0}^{2\pi /\Lambda }k\, dk\int _{0}^{\pi }d\theta  \).
After performing the trivial integration over \( dk \) we have
\begin{equation}
F  =  d +\frac{T'}{2}\left[I-\pi \ln \left(
-\frac{\Lambda ^{D}T'}{\ell ^{D}} \right) \right]
\label{F1}
\end{equation}
where $I$ is the integral
\begin{equation}
I  =  \int ^{\pi /2}_{-\pi /2}\ln (-w^{+})d\theta.
\label{I}
\end{equation}
The  integrand $w^+$ is a temperature-independent strain function
so that $-\partial F/\partial T'$ gives 
\begin{equation}
s=-\frac{I}{2} + \frac{\pi}{2} \left[1 + \ln \left(-\frac{\Lambda^D T'}{\ell^D}\right)\right]
\label{s1}
\end{equation}
while  from $F=U-T's$
\begin{equation}
U=d + \frac{\pi}{2} T'.
\label{uresult}
\end{equation}
Since \(d\) represents the linear elastic response of an intact rock, and \( T' \) 
decreases from  zero to negative values as damage accumulates, 
this expression shows that the average energy  decreases due to  the presence of  cracks 
and is thus consistent with the  negative curvature of the strain/stress load curve observed 
experimentally.

Before addressing how  $s$  and $F$ (and their derivatives)  behave  at localization, we
first establish the
stress and temperature behavior at localization.

\subsection{Mechanical behavior  at localization}
Consider  the stress components
$\sigma = -2 {\partial F}/{\partial \gamma}$ and  
$p = -2 {\partial F}/{\partial \Delta}
$
where $\sigma$ (shear stress) and $p$ (pressure) are
both positive and related to the axial and radial stress components as
\begin{equation}
-\sigma   =  \tau _{a}-\tau_{r}, \quad \mbox{ and } \quad
-p   =  \tau _{a} + \tau _{r}.
\label{stressdefs}
\end{equation}
In standard lab experiments, the axial stress $\tau_a$ varies while
the radial stress $\tau_r = -p_r $  is kept constant. 
  The
strain components   $\gamma$ (shear strain) and $\Delta$ (dilatation)
 are similarly related to  the axial and radial strain  as
\begin{equation}
\gamma   =  \varepsilon _{a}-\varepsilon _{r}, \quad \mbox{ and } \quad
\Delta   =  \varepsilon _{a}+\varepsilon _{r}.
\label{straindefs}
\end{equation}
Using the definition of $w^+$ [Eq.\ (\ref{w+definition})]
along with the definitions of  $L_{\ka}$, $M_{\ka}$ and
$N_{\ka}$ given in the summary of Paper II, we  differentiate
the integral $I$ with respect to the actual strain variables,
evaluate along the load path ($\Delta_m = \Delta$ and $\gamma_m = \gamma$),
 use the definition $\omega = \omega_3 \gamma/ \Delta$ with the new constant $\omega_3 = \kappa_3 / \kappa_1 $ and make
the change of integration variables $z = \tan^{-1}\theta$ to obtain exactly
\begin{eqnarray}
\partial _{\gamma }I & = & \frac{\omega _{3}}{(1-q) \Delta }{\partial_\omega I}
\label{dIdgamma} \\
\partial _{\Delta }I & = & \frac{1}{(1-q) \Delta } \left( 2\pi - \omega
{\partial_\omega I} \right) \label{dIddelta}
\end{eqnarray}
where $ q=1-1/(k+2)$ is the constant associated with the exponent $k\ge0$
 of the quenched disorder distribution, and the integral $\partial_\omega I$  is defined
\begin{equation}
\partial_\omega I
 =  \int ^{+\infty }_{-\infty }\frac{\partial _{\omega }g}{g}\frac{dz}{1+z^{2}}\label{J}
\end{equation}
with $g(\omega,z)$  given by
\begin{eqnarray}
g(\omega ,z) & = &
 [1-\alpha -2(1-\alpha )c+(1-\alpha )c^{2}+\omega ^{2}]z^{4} \nonumber \\
&  & +[4\alpha \omega +4(1-\alpha )c\omega] z^{3}\nonumber \\
&  & +[2+2\alpha +2(1-\alpha )c^{2}-2(2\alpha -1)\omega ^{2}]z^{2}\nonumber \\
&  & +[-4\alpha \omega +4(1-\alpha )c\omega ]z \nonumber \\
& &  + 1-\alpha +2(1-\alpha )c+(1-\alpha )c^{2}+\omega ^{2}.
 \label{defg}
\end{eqnarray}
  Thus, the shear stress and pressure can be written
\begin{eqnarray}
\sigma  & = & -2(1-\alpha )\gamma -\frac{T'}{\Delta}\frac{\omega_{3}}{(1-q) }
\partial _{\omega }I(\omega )\label{sigma} \\
-p & = & 2\alpha \Delta +\frac{T'}{(1-q) \Delta}\left[ {2\pi }-
{\omega}
\partial _{\omega }I(\omega )\right]. \label{p}
\end{eqnarray}
The integral $\partial_\omega I$ is  solved using the residue theorem once the roots $z$ of
the quartic $g(\omega,z)$ have been found.

This quartic decomposes into the exact  form
\begin{eqnarray}
g(\omega ,z) & = & [z-\zeta (\omega )][z-{\zeta }^*(\omega )]u(\omega ,z) 
\label{g2}\\
u(\omega ,z) & = & \rho (\omega )[z-\xi (\omega )][z-{\xi }^*(\omega )]
\label{g3}
\end{eqnarray}
where the star indicates taking the complex conjugate.  The roots
\( \zeta (\omega ) \) and $\zeta^*(\omega)$ both merge to the real axis
in the approach to localization $\omega \rightarrow \omega_c$, while
the other two roots
 \( \xi (\omega ) \) and $\xi^*(\omega)$  remain complex at
localization.

There are thus  three simple poles \( \zeta (\omega ) \), \( \xi (\omega ) \)
and \( i \) contributing to $\partial_\omega I$ if the loop is closed in the  upper-half
$z$ plane so that  the residue theorem yields
\begin{eqnarray}
\frac{\partial_\omega I}{\pi }&=&
\frac{\partial _{\omega }g(\zeta )}{\Im \{\zeta \} u(\zeta)[1+\zeta ^{2}]}
+\frac{\partial _{\omega }g(\xi )}{\rho \Im \{\xi \}
[\xi -\zeta ][\xi -{\zeta }^*][1+\xi ^{2}] } \nonumber \\
&&+\frac{\partial _{\omega }g(i)}{[i-\zeta ][i-{\zeta }^*]u(i)}
\label{Iomegafirst}
\end{eqnarray}
where \( \Im \{ \,  \} \) designates taking  the imaginary part.
We are interested in evaluating this integral (and therefore the roots $\zeta$ and
$\xi$ and the function $\rho$)
only in the approach to localization; {\em i.e.}, when $\delta \omega = \omega - \omega_c$
can be considered small.  In this limit,  the second and third terms of Eq.\ (\ref
{Iomegafirst})  (the residues
from $\xi$ and $i$) have numerators
and denominators that are both order 0 in $\delta \omega$ so that
it suffices to know the behavior
\begin{eqnarray}
\xi (\omega ) & = & \xi _{0}+\xi _{1}\delta \omega \\
\rho (\omega ) & = & \rho _{0}+\rho _{1}\delta \omega.
\end{eqnarray}
However, the residue related to $\zeta$ is proportional to  $\delta \omega$  in both 
the numerator and denominator which requires knowledge of this root to second order
\begin{equation}
\zeta (\omega )  =  \zeta _{0}+\zeta _{1}\delta \omega +\zeta _{2}\delta \omega ^{2}.
\end{equation}
The various strain-independent
constants $\xi_i$, $\rho_i$, and $\zeta_i$ are all known groupings of the
elastic constants derived from Eqs.\ (\ref{defg}), (\ref{g2}) and (\ref{g3}).  
The final result for the integral after an enormous algebraic reduction is
\begin{equation}
\partial_\omega I =  I_c + I_1 \delta \omega
\end{equation}
where the constants $I_c$ and $I_1$ are exactly
\begin{equation}
I_c = 2 \pi \frac{\sqrt{c^2 -1}}{c^2} \quad \mbox{ and } \quad
I_1 = 2 \pi \frac{2-c^2}{c^4}.
\label{IcI1}
\end{equation}

\subsubsection{Stress and strain at localization}
The shear stress and pressure may be written
$$\sigma = \sigma_0 + \sigma^{\rm int} \quad \mbox{ and }
p=p_0 + p^{\rm int}$$
where $\sigma_0=-2(1-\alpha)\gamma$ and $p_0= -2 \alpha \Delta $
are the trivial linear
variations of the uncracked material.  We have just shown that
at localization ($\delta \omega =0$), the non-trivial shear stress due
to cracks and crack interaction is
\begin{equation}
\sigma_c^{\rm int} = - \frac{2\pi \omega_3 T'_c}{(1-q) \Delta_c}
\frac{\sqrt{c^2-1}}{c^2} < 0
\label{scint}
\end{equation}
while the non-trivial pressure  is
\begin{equation}
p_c^{\rm int} = - \frac{2\pi T'_c}{(1-q) \Delta_c c^2} < 0.
\label{pcint}
\end{equation}
That these critical values are both negative follows because 
$T_c'$ (scaled temperature at localization) is negative  and  
 $\Delta_c$ (total dilatation at localization) will soon be shown to be negative. 
 Equations (\ref{scint}) and (\ref{pcint})  say that the presence of cracks has
lowered both shear stress and pressure relative to an intact material
at the same strain.  This is indeed what is observed in experiments.  

To quantify  the nature of $\Delta_c$, we use  that
the confining pressure $p_r$ is a known positive
constant in standard  experiments on
rocks so that  
\begin{equation}
\label{implicit,eq}
p_{r}=-\alpha \Delta _{c}+(1-\alpha )\gamma _{c}
-\frac{T_{c}'[2\pi -(\omega _{c}+\omega_{3})I_{c}]}{2 (1-q) \Delta _{c}}.
\end{equation}
Together with $\omega_c = \omega_3 \gamma_c/ \Delta_c$, this  represents  
 an equation for $\Delta_c$
\begin{equation}
\label{eqforroot}
\!\!\left[\alpha -(1-\alpha )\frac{\omega_{c}}{\omega _{3}}\right] \!\!
\Delta _{c}^{2}+p_{r}\Delta _{c}+\frac{T_{c}'\left[2\pi -(\omega _{c}
+\omega_{3}) I_{c}\right]}{(1-q)}\!\! =0.
\end{equation}
Because $T'$ varies with strain, we have that $T'_c$ is also a function
of $\Delta_c$  so that  Eq.\ (\ref{eqforroot}) is more than a  simple  quadratic in $\Delta_c$.
To obtain an order-of-magnitude estimate of  $T'_c$, we use the approximate temperature
expression based on non-interacting cracks 
\begin{eqnarray}
\!\!\!\!\!\!\!\!\!\!\!\!\!\!
&&\frac{1}{T'_c}=-\frac{2 \Lambda^2}{d_m^2 (1-q)[\kappa_2 + \kappa_3 (\omega_c/\omega_3)^2] \Delta_c^2}
\nonumber \\
\!\!\!\!\!\!\!\!\!\!\!\!\!
&&\times    \ln\left\{ \left[\frac{2\Gamma }{(\lambda +2\mu )
d_{m}\Delta_c^2 [\kappa _{2}+\kappa_3 (\omega_c/\omega_{3})^2]}\right]^{q/(1-q)}
\!\!\!\!\!\!-1\right\}.
\label{approxtemp}
\end{eqnarray}
 After putting Eq.\ (\ref{approxtemp}) into  Eq.\ (\ref{eqforroot}), $\Delta_c$ is 
numerically
determined  using Newton's method. The predicted 
$\Delta_c$ is negative for the range of confining pressure $p_r$ of interest and 
 remains negative for all 
 ranges of elastic moduli found in rocks. 
 The signs of the various terms in Eq.\ (\ref{eqforroot}) imply that the 
transition happens when the temperature has sufficiently departed from zero, 
but is still negative.  Typical results from the numerical evaluation are 
$T_{c}' \sim - 10^{-2} (\lambda + 2 \mu)$, 
which confirms the rough  estimate given in Sec.\ V of Paper II. The typical value for 
$\Delta_c$ is  a few percent;  {\em i.e.},  the order of magnitude  experimentally observed  at peak stress.

The conclusion is that at localization,  both dilatation $\Delta_c$ and
shear strain $\gamma_c = \omega_c \Delta_c /\omega_3$  are negative while    
$|\gamma_c| \gg |\Delta_c|$.  This demonstrates that 
 the radial strain $\varepsilon_{r}=\Delta_c-\gamma_c$ is positive at localization, 
which is also  consistent with experimental observations.  

\subsubsection{Stress, strain, and temperature derivatives at localization}
We now address how the stress and strain components as well as the
temperature are changing  with the negative of axial strain $\varepsilon = -\varepsilon_a =
-(\Delta + \gamma)/2$
at localization.

In the approach to localization we write $ \Delta = \Delta_c +\delta \Delta$, \
$\gamma = \gamma_c + \delta \gamma$, and $T' = T'_c + \delta T'$  using 
the exact differential equation for temperature to define $\delta T'$ in what follows 
(not the
 approximation).   The condition that $p_r$ is
constant requires that  
\begin{eqnarray*}
 &  & \delta \Delta \left\{-\alpha +\frac{T_{c}'\left[2\pi -(2\omega _{c}+\omega_{3})I_{c}
-(\omega _{c}+\omega_{3})\omega _{c}I_{1}\right]}{2(1-q)\Delta ^{2}_{c}}
\right\}\\
 &  & +\delta \gamma \left\{1-\alpha +\frac{T_{c}'\omega_{3}\left[I_{c}+(\omega _{c}
+\omega_{3})I_{1}\right]}{2(1-q)\Delta ^{2}_{c}} 
 \right\}\\
 &  & -\frac{\delta T'}{2(1-q)\Delta _{c}}\left\{2\pi
-(\omega _{c}+\omega_{3})I_{c}\right\} = 0
\end{eqnarray*}
which along with
$-2\delta \varepsilon   =  \delta \Delta +\delta \gamma$
 gives
\begin{widetext}
\begin{eqnarray}
\frac{1}{2}\frac{d\Delta}{d \varepsilon}  & = &
\frac{ 2(1-q)(1-\alpha)\Delta_c^2/T'_c +\omega_{3} 
\left[I_{c} +(\omega _{c}+\omega_{3})I_{1}\right]
+{\Delta _{c}}/(2T'_c) \left[ 2\pi -(\omega _{c}
+\omega_{3})I_{c}\right] dT' / d \varepsilon } 
{-2(1-q)\Delta_c^2/{T_{c}'}+ 
 2\pi -2(\omega _{c}+\omega_{3})I_{c}-(\omega _{c}
+\omega_{3})^{2}I_{1} } \label{ddeltadepsT} \\
\frac{1}{2}\frac{d \gamma}{d\varepsilon}  & = & \frac{ 2(1-q)\alpha \Delta_c^2/{T_{c}'} - 
2\pi +(2\omega _{c}+\omega_{3})I_{c}+(\omega _{c}
+\omega_3) \omega_{c} I_{1} 
-{\Delta _{c}}/(2T'_c) \left[ 2\pi -(\omega _{c}
+\omega_{3})I_{c}\right] dT' / d \varepsilon}{-2(1-q)\Delta_c^2/{T_{c}'} + 
 2\pi -2(\omega _{c}+\omega_{3})I_{c}-(\omega _{c}
+\omega_{3})^{2}I_{1} } .\label{dgammadepsT}
\end{eqnarray}
To obtain  an exact
 expression for $dT'/d\varepsilon$ (within the context of having employed
the mean-phase approximation), we use the formalism
of Sec.\ IV.\ A of Paper I 
to write 
\begin{eqnarray}
\left[ \partial _{T'}U+\left(\partial _{\Delta }U+\partial _{\Delta _{m}}U+\frac{p}{2}\right)
\partial _{T'}\Delta
 +\left(\partial _{\gamma }U+\partial _{\gamma _{m}}U
+\frac{\sigma }{2}\right)\partial _{T'} \gamma \right] \frac{dT'}{d\varepsilon } &&\nonumber \\
+\left(\partial _{\Delta }U+\partial _{\Delta _{m}}U+\frac{p}{2}\right)
\partial _{\varepsilon }\Delta
+\left(\partial _{\gamma }U+\partial _{\gamma _{m}}U
+\frac{\sigma }{2}\right) \partial _{\varepsilon }\gamma  & &=  0. \label{temp,d.e.}
\end{eqnarray}
Using Eq.\ (\ref{uresult}) for $U$, we  have
$\partial _{T'}U  =  {\pi }/{2}$,
$\partial _{\Delta }U  =  \alpha \Delta =-p_{0}/2$,
$\partial _{\gamma }U  =  (1-\alpha )\gamma =-\sigma _{0}/2$,
$\partial _{\Delta _{m}}U  =  0$, and
$\partial _{\gamma _{m}}U  =  0$
so that  the temperature derivative
at localization is given by
\begin{equation}
\frac{1}{2}\frac{dT'}{d\varepsilon }=\frac{\left[(2\pi- \omega_{c} I_{c}) (1-\alpha )
+\omega_{3}I_{c} \alpha \right]2(1-q) \Delta_c 
+ \omega_{3}
(\omega _{c}+\omega_{3})(2\pi I_1
+\omega_{3} I_{c}^{2}) T'_c / \Delta_c  }
{-[2\pi -(\omega _{c}
+\omega_{3})I_{c}]^{2}
+(1-q)\pi
\left[-2(1-q)\Delta_c^2/T'_c +2\pi -2(\omega _{c}
+\omega_{3})I_{c}-(\omega _{c}
+\omega_{3})^{2}I_{1}\right] }.
\end{equation}
\end{widetext}
This  derivative
is numerically calculated to be finite and negative
for the ranges of elastic moduli and radial confining
pressures of interest, thus indicating that the localization transition
always preceeds the phase transition where the temperature diverges to $-\infty$.
Since rocks fail immediately after localization, the temperature-divergence 
transition is not observed in rock experiments.

Last, we determine the variation of the stress components with axial strain $\varepsilon$
at localization. Since $p_r$ is constant, we have that  $dp/d\varepsilon=d\sigma/d\varepsilon=
-d\tau_a/d\varepsilon$.  These derivatives   define the so-called ``tangent modulus''  given by
\begin{eqnarray}
\frac{1}{2}  \frac{d \sigma }{d\varepsilon} & = & 
-\frac{\omega_{3}}{2(1-q)}\frac{I_{c}}{\Delta _{c}} \frac{dT'}{d \varepsilon} +
\frac{T_{c}'\omega_{3}(I_{c}+\omega _{c}I_{1})}{2(1-q)\Delta ^{2}_{c}}\frac{d\Delta}{d\varepsilon} \nonumber \\
 &  & -\left[ (1-\alpha )+\frac{T_{c}'}{2(1-q)\Delta ^{2}_{c}}\omega_{3}^{2}I_{1}\right] \frac{d \gamma}{d\varepsilon} 
\end{eqnarray}
where the derivatives $d\Delta/d\varepsilon$, $d\gamma/d\varepsilon$, and $dT'/d\varepsilon$
have been given above.

\begin{figure}
{\par\centering \resizebox*{8cm}{7cm}{\includegraphics{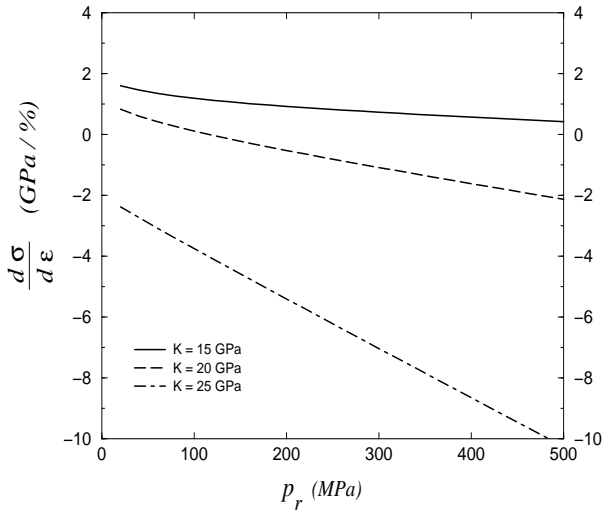}} \par}
\caption{ The localization value  of the axial tangent modulus 
$d\sigma/ d \varepsilon $ 
as a function of the radial confining pressure $p_r$. The three curves represent 
different assumed bulk  moduli for the mineral.   The other rock properties are $\Gamma = 10 $ J/m$^2$, 
$d_m = 10 \mu$m, $\mu=15$ GPa, and $q=3/4$. } 
\label{stressfig}
\end{figure}

In Fig.\ \ref{stressfig}, we plot how  $d\sigma/d \varepsilon$  
 varies with radial confining pressure for various values 
of the elastic constants.  The plot shows that for a sufficiently large  ratio of  bulk to shear modulus, 
the axial pressure is  always decreasing at 
localization, which means that it has already passed through the stress maximum. 
However, for sufficiently small bulk moduli and at low confining pressures, 
localization can also occur prior to peak stress.  Thus,   
 peak stress and localization are distinct in our theory.   Localization 
can occur in either the hardening or softening regime depending on the bulk modulus and confining pressure. 
 When localization  occurs in the softening regime (large bulk modulus), 
the strain/stress curve around peak stress is necessarily an analytic (quadratic) function, 
whereas when it occurs in the hardening regime (small bulk moldulus with small confining pressure),  
the peak stress presumably corresponds  
to a sharper variation  as  microcracks start to coalesce along a  weakened band and 
unstable failure sets in. These predictions are consistent with the experimental observations.

\subsection{Entropy and its derivatives at localization}
The exact result  $\partial_\omega I = I_c + I_1 \delta\omega$ with $I_c$ and $I_1$ 
as given by Eq.\ (\ref{IcI1})  
 means that the integral $I$ of Eq.\ (\ref{I}) 
is itself both finite 
and continuous in the limit as $\delta \omega \rightarrow 0$.  Because it 
has further been shown that $T'$ remains finite and continuous at localization, 
 Eqs.\ (\ref{F1}) and  (\ref{s1}) then show that both the free energy and 
the entropy (and all of their derivatives with respect to strain)  
remain finite and continuous as $\delta \omega \rightarrow 0$.  This demonstrates 
 exactly that the localization transition is a continuous 
phase transition and allows us  to classify it as a critical point.

\section{Correlation function\label{sect:correl,funct}}

\subsection{Derivation of a diverging correlation length}

The qualitative study of Sec.\ \ref{sect:presence,of,transition} lead to the
conclusion that the localization transition is associated with the creation
of conjugate bands of coherently oriented cracks. In this final section, the statistical
correlation between cracks will be quantitatively addressed.

The autocorrelation function is defined 
\begin{equation}
\label{eq:def:autocorrel}
G(\iks ,\igr )=\left\langle \varphi (\iks )\varphi (\igr )\right\rangle -\left\langle \varphi (\iks )\right\rangle \left\langle \varphi (\igr )\right\rangle 
\end{equation}
 and will be determined  using a standard method of statistical mechanics
\cite{LB90b,Gol92,BDF+92}. First, the Hamiltonian \( E[\varphi ] \) is generalized
to include an additional coupling of the local field \( \varphi (\iks ) \)
with an aribitrary field \( J(\iks ) \) coming from some external source 
\begin{equation}
\label{eq:hamilton, generalise}
E'[\varphi ,J]=E[\varphi ]-\int _{\iks \in \Omega }d^{D}\iks \, J(\iks )\varphi (\iks ).
\end{equation}
 The partition function becomes then a functional of the external field 
\begin{equation}
\label{eq:Z,general}
Z[J]=\int \underset {\iks \in \Omega }{\prod }(d\varphi _{\iks })e^{-E'[\varphi ,J]/T}
\end{equation}
 and the averages involved in Eq.\ (\ref{eq:def:autocorrel}) are obtained by
taking functional derivatives of \( Z[J] \) with respect to \( J \) and then
letting the external field go to zero; \emph{i.e.}, 
\begin{eqnarray}
\left\langle \varphi (\iks )\right\rangle  & = & \lim _{J\rightarrow 0}\frac{T}{Z}\frac{\delta Z}{\delta J(\iks )}\label{eq:meanphi} \\
\left\langle \varphi (\iks )\varphi (\igr )\right\rangle  & = & \lim _{J\rightarrow 0}\frac{T^{2}}{Z}\frac{\delta ^{2}Z}{\delta J(\iks )\delta J(\igr )}.\label{eq:meanphiphi} 
\end{eqnarray}
 Since the original Hamiltonian is most easily handled in Fourier form, the
external coupling will be expressed as 
\begin{eqnarray}
\lefteqn {-\int _{\iks \in \Omega }d^{D}\iks \, J(\iks )\varphi (\iks )=} &  & \nonumber \\
 &  & -\frac{1}{\ell ^{D}}\left( \widetilde{J}_{0}\widetilde{\varphi }_{0}+2\sum _{\ka \in \Upsilon }\widetilde{J}^{R}_{\ka }\widetilde{\varphi }^{R}_{\ka }+2\sum _{\ka \in \Upsilon }\widetilde{J}^{I}_{\ka }\widetilde{\varphi }^{I}_{\ka }\right) \label{eq:Fourier,ext,coupling} 
\end{eqnarray}
 where the superscripts \( R \) and \( I \) refer once again to the real and
imaginary part of a complex quantity. The functional derivatives relative to
\( J(\iks ) \) must then be expressed by their counterparts in Fourier space
\begin{eqnarray}
\frac{\delta }{\delta J(\iks )}\! \!  & = & \! \! \! \! \! \sum _{\ka \in \Upsilon \cup \{0\}}\! \! \left( \frac{\delta \widetilde{J}^{R}_{\ka }}{\delta J(\iks )}\frac{\partial }{\partial \widetilde{J}^{R}_{\ka }}+\frac{\delta \widetilde{J}^{I}_{\ka }}{\delta J(\iks )}\frac{\partial }{\partial \widetilde{J}^{I}_{\ka }}\right) \nonumber \label{eq:tfdel} \\
 & = & \! \! \! \! \! \sum _{\ka \in \Upsilon \cup \{0\}}\! \! \! \left( \! \cos (\ka \pd \iks )\frac{\partial }{\partial \widetilde{J}^{R}_{\ka }}-\sin (\ka \pd \iks )\frac{\partial }{\partial \widetilde{J}^{I}_{\ka }}\right) \! .\label{eq:tfdelta} 
\end{eqnarray}
 The modified partition function will again be determined using the constant-sign
approximation, but now the presence of the external field breaks the symmetry
between the sum over positive and negative crack-fields, so that both terms
need to be kept in the generalization of Eq.\ (\ref{eq:Z,negatifs}). This leads
to a slightly more complicated version of Eq.\ (\ref{eq:Z,reduced}) for the
expression of \( Z \) in the thermodynamic limit 
\begin{equation}
\label{eq:Z,general,reduced}
Z\simeq z_{0}\left\{ z^{+}_{1}\underset {\ka \in \Upsilon }{\prod }\left[ z_{R}^{+}(\ka )z_{I}^{+}(\ka )\right] +z^{-}_{1}\underset {\ka \in \Upsilon }{\prod }\left[ z_{R}^{-}(\ka )z_{I}^{-}(\ka )\right] \right\} 
\end{equation}
 where \( z_{0} \) is again the trivial intact term,  and where
\begin{eqnarray}
z^{\pm }_{1} & = & \int _{x\in \mathbb {R}^{+}}dx\, e^{-\left( e\ell ^{D}\pm \widetilde{J}_{0}\right) x]/\Lambda ^{D}T'}\label{eq:z0pm} \\
z_{R}^{\pm }(\ka ) & = & \int _{x\in \mathbb {R}}dx\, e^{-\left[ -2\widetilde{J}^{R}_{\ka }x+\ell ^{D}w^{\pm }(\ka )x^{2}\right] /\Lambda ^{D}T'}\label{eq:zpm} 
\end{eqnarray}
 with \( z_{I}^{\pm }(\ka ) \) having the  same form as   \( z_{R}^{\pm }(\ka ) \)
after replacing \( \widetilde{J}^{R}_{\ka } \) with \( \widetilde{J}^{I}_{\ka } \).
In the following, the forms implying derivatives with respect to \( \widetilde{J}^{I}_{\ka } \)
are to be implicitly understood as having the same form as their counterpart
with respect to \( \widetilde{J}^{R}_{\ka } \) (this imaginary components will
not be explicitly written out).

The integrals are easily performed giving 
\begin{eqnarray}
 &  & z_{1}^{\pm }=\Lambda ^{D}T'/\left[ e\ell ^{D}\pm \widetilde{J}_{0}\right] \label{eq:z0plus,asympt} \\
 &  & z_{R}^{\pm }(\ka )=\exp \left[ \frac{\left( \widetilde{J}^{R}_{\ka }\right) ^{2}}{\Lambda ^{D}\ell ^{D}w^{\pm }(\ka )T'}\right] \sqrt{\frac{\pi \Lambda ^{D}T'}{\ell ^{D}w^{\pm }(\ka )}}.\label{eq:zkaplus,asympt} 
\end{eqnarray}
 The first derivatives of \( Z \) with respect to the external field are then
\begin{eqnarray}
\frac{\partial Z}{\partial \widetilde{J}_{0}} & \! \! = & \! \! z_{0}\left\{ -\frac{\Lambda ^{D}T'}{(e\ell ^{D}+\widetilde{J}_{0})^{2}}z^{+}_{1}\underset {\ka \in \Upsilon }{\prod }\left[ z_{R}^{+}(\ka )z_{I}^{+}(\ka )\right] \right. \nonumber \\
 & \! \!  & \! \! \quad \quad \left. +\frac{\Lambda ^{D}T'}{(e\ell ^{D}-\widetilde{J}_{0})^{2}}z^{-}_{1}\underset {\ka \in \Upsilon }{\prod }\left[ z_{R}^{-}(\ka )z_{I}^{-}(\ka )\right] \right\} \label{eq:d1ZJ0} \\
\frac{\partial Z}{\partial \widetilde{J}^{R}_{\ka }} & \! \! = & \! \! z_{0}\left\{ z^{+}_{1}\underset {\ka \in \Upsilon }{\prod }\frac{2\widetilde{J}^{R}_{\ka }}{\Lambda ^{D}\ell ^{D}w^{+}(\ka )T'}\left[ z_{R}^{+}(\ka )z_{I}^{+}(\ka )\right] \right. \nonumber \\
 & \! \!  & \! \! \quad  \left. +z^{+}_{1}\underset {\ka \in \Upsilon }{\prod }
\frac{2\widetilde{J}^{R}_{\ka }}{\Lambda ^{D}\ell ^{D}w^{-}(\ka )T'}
\left[ z_{R}^{-}(\ka )z_{I}^{-}(\ka )\right] \right\} .\label{eq:d1ZJk} 
\end{eqnarray}
 Letting the external field go to zero, both of these terms disappear, so that
using the chain rule of Eq.\ (\ref{eq:tfdelta}), the average of the crack variable
\( \varphi  \) at any point \( \iks  \) in a mesovolume is given by Eq.\ (\ref{eq:meanphi})
to be 
\[
\left\langle \varphi (\iks )\right\rangle =0.\]
 As expected, there is no spontaneous symmetry breaking prior to the transition.

Consequently, the autocorrelation function reduces to only the second derivatives
of \( Z \) in Eq.\ (\ref{eq:meanphiphi}). Differentiating Eqs.\ (\ref{eq:d1ZJ0})--(\ref{eq:d1ZJk})
with respect to \( \widetilde{J}_{0} \), \( \widetilde{J}^{R}_{\ka '} \) and
\( \widetilde{J}^{I}_{\ka '} \), and taking the limit where \( J \) goes uniformly
to zero leads to 
\begin{eqnarray*}
\!\!\!\!\frac{\partial ^{2}Z}{\partial \widetilde{J}^{R}_{\ka }\partial \widetilde{J}^{R}_{\ka '}} 
& \!\!= \!\!& 
\frac{\partial ^{2}Z}{\partial \widetilde{J}^{I}_{\ka }\partial \widetilde{J}^{I}_{\ka '}}
= \left[ \frac{1}{w^{+}(\ka )}+\frac{1}{w^{-}(\ka )}\right] \!\!
\frac{2 Z_0 \delta _{\ka \ka '}}{\Lambda ^{D}\ell ^{D}T'} \\
\frac{\partial ^{2}Z}{\partial \widetilde{J}^{2}_{0}} & = & 
\frac{2\Lambda ^{D}T'}{\ell ^{3D}e^{3}}Z_{0} 
\end{eqnarray*}
 where \( Z_{0}=Z[0] \) is the original partition function without external
source.   All the remaining cross-derivatives go to zero 
\begin{equation*}
 \frac{\partial ^{2}Z}{\partial \widetilde{J}_{\ka }^{R}\partial \widetilde{J}^{I}_{\ka '}}
=\frac{\partial ^{2}Z}{\partial \widetilde{J}_{\ka }^{R}\partial \widetilde{J}_{0}}
=\frac{\partial ^{2}Z}{\partial \widetilde{J}_{\ka }^{I}\partial \widetilde{J}_{0}}=0.
\end{equation*}
 Through the chain rules of Eq.\ (\ref{eq:tfdelta}), these equalities show
that the autocorrelation function has the form \( G(\iks ;\igr )=G(\iks -\igr ) \)
due to the symmetry of the problem under translation for an infinite system.
The Fourier transform  \( G(\iks ;\igr )
=\sum _{\ka }\widetilde{G}_{\ka }e^{i\ka \cdot (\iks -\igr )}/\ell ^{D} \)
is thus given by 
\begin{equation}
\label{eq:TFG}
\widetilde{G}_{\ka }=2\Lambda ^{D}T'\left[ \frac{1}{w^{+}(\ka )}+\frac{1}{w^{-}(\ka )}\right] 
\end{equation}
 when \( \ka \neq 0 \). The special value \( \widetilde{G}_{0}=2\Lambda ^{3D}{T'}^{3}/e^{3}\ell ^{3D} \)
does not play any role in the thermodynamic limit.

In real space, the autocorrelation function is obtained by an inverse Fourier
transform 
\begin{equation}
\label{eq:TFGprinciple}
G(\iks )=\frac{1}{4\pi ^{2}}\int ^{2\pi /\Lambda }_{0}kdk\int ^{2\pi }_{0}d\theta \widetilde{G}(\ka )e^{i\ka \cdot \iks }.
\end{equation}
Using \( \widetilde{G}(-\ka )=\widetilde{G}(\ka ) \) which is a consequence
of \( w^{\pm } \) being \( \pi  \)-periodic functions, and working in polar-coordinates
\( \iks =(x,\theta _{x}) \) and \( \ka =(k,\theta ) \), the angular integral
is  divided into two symmetric domains which  gives  
\[
G(x,\theta _{x})=\!\!\int ^{2\pi /\Lambda }_{0}\!  \frac{kdk}{2\pi ^{2}}\int ^{\theta _{x}+\pi }_{\theta _{x}}
\! \! \! d\theta \widetilde{G}(k,\theta )\cos \left[ kx\cos \left( \theta -\theta _{x}\right) \right] \]
Since \( w^{\pm } \) and therefore \( \widetilde{G} \)  only depend on the
angular part of \( \ka  \), the integral over \( k=|\ka|  \) yields   
\begin{eqnarray*}
\lefteqn {G(x,\theta _{x})=\int ^{\theta _{x}+\pi }_{\theta =\theta _{x}}d\theta \, 
 \frac{\widetilde{G}(\theta )}{2\pi ^{2}}\pd } &  & \\
 &  & \times \left[ \frac{2\pi }{\Lambda x\cos (\theta -\theta _{x})}\sin 
\left( \frac{2\pi x\cos (\theta -\theta _{x})}{\Lambda }\right) \right. \\
 &  & \left. +\frac{1}{x^{2}\cos ^{2}(\theta -\theta _{x})}
\left\{ \cos \left( \frac{2\pi x\cos (\theta -\theta _{x})}{\Lambda }\right) -1\right\} \right].  
\end{eqnarray*}
For \( x\gg \Lambda  \), this integral is dominated by a neighborhood
of \( \theta =\theta _{x}+\pi /2 \), of angular size \( c_{1}\Lambda /x \)
with \( c_{1} \) a constant of order unity. The function \( \widetilde{G}(\theta ) \)
is almost constant over this small neighborhood, and this integral can be well
approximated as 
\[
G(x,\theta _{x})=\frac{\widetilde{G}(\theta _{x}+\pi /2)}{\Lambda ^{2}}
I_G \left(\frac{x}{\Lambda }\right)  
\]
with the dimensionless integral \( I_G \) defined as
\begin{eqnarray*} 
I_G(u)&=& 2 \pi^2\int ^{\pi }_{\theta =0}d\theta \,    
 \left[ \frac{2\pi \sin \left( 2\pi u\cos (\theta )\right) }
{u\cos (\theta )} \right. \\
 &  & \qquad \qquad 
\left. +\frac{\left\{ \cos \left( 2\pi u\cos (\theta )\right) -1\right\} }{u^{2}\cos ^{2}(\theta )}
 \right].   
\end{eqnarray*}
 An asymptotic study of this oscillating integral for \( u\gg 1 \) shows that
\( I_G(u)\sim c_{2}/u \), with \( c_{2} \) a positive constant of order unity.
Reformulating Eq. (\ref{eq:TFG}) with 
\begin{equation}
\label{def,h}
h(\theta )=2c_{2}T'\left [\frac{1}{w^{+}(\theta +\pi /2)}+\frac{1}{w^{-}(\theta +\pi /2)}\right ]
\end{equation}
gives  the  real space  autocorrelation function 
in the form 
\begin{equation}
\label{result,autocorrel}
G(x,\theta _{x})\sim h(\theta _{x})\frac{\Lambda }{x}.
\end{equation}
This establishes  that along any direction,  the autocorrelations decay as
\( \Lambda /x \) (for two points  separated by a significant
number of grains, \( x\gg \Lambda  \)). 

Concerning the  angular dependence of $G$,  the symmetry
of the system under parity leads to \( w^{-}(\theta )=w^{+}(\pi /2-\theta ) \)
(which can also be verified directly from the definitions of \( w^{\pm } \)
and the dependancies of \( L_{\ka },M_{\ka },N_{\ka } \) on \( u_{\ka }=\cos (2\theta ) \)
and \( v_{\ka }=\sin (2\theta ) \) given in paper II, together with the fact that
the parity symmetry keeps \( v \) constant but changes the sign of \( u \)).
This, along with  the \( \pi  \) periodicity of \( w^{\pm } \),  shows that  
$G$ is  symmetric under parity;  {\em i.e.},   \( G(x,\pi /2-\theta _{x})=G(x,\theta _{x}) \). 

The angular dependence is  best shown  by considering curves of iso-correlations 
 \( G(x,\theta _{x})=c_{3} \) where $c_3$ is constant along
a  curve. Such  curves obey \( x=\Lambda h(\theta _{x})/c_{3} \). The direct study of the function \( w^{+} \) shows that it admits quadratic maxima along the directions \( \theta ^{+}[\pi ] \),
scaling as \( \max (w^{+})=w^{+}(\theta ^{+}[\pi ])=-a (\delta \omega)^{2} \) when the transition
is approached, where $a$ is a positive constant. This comes from the fact that $E^{\text{int}}$ is degenerate exclusively for the critical angles $ \theta ^{\pm}$, at reduced strain $\omega_c$. Outside a small neighborhood of \( \theta ^{+}[\pi ] \), \( w^{+} \)
remains bounded. The definition of \( h \) and the exchange under parity of
\( w^{\pm } \) shows then that such an iso-correlation curve has  four branches
(spikes) along the directions \( \pm \theta ^{+}\pm \pi /2 \), whose extent
\( \xi  \) diverge to \( +\infty  \) as 
\begin{equation}
\xi \sim 2\Lambda (-T')\frac{c_{2}}{a c_{3}}(\omega_c-\omega)^{-2}
\sim c_{4}\Lambda (\varepsilon _{c}-\varepsilon )^{-2}.
\end{equation}
 The fact that \( w^{+} \) remains bounded outside any small neighborhood of
\( \theta ^{+}[\pi ] \) also means that the width \( \rho  \) of the branches
remains finite;  {\em i.e.},  that the aspect ratio of the branches \( \xi /\rho  \)
also diverges  as \( (\varepsilon _{c}-\varepsilon )^{-2} \). This is qualitatively
illustrated in Figure 3.
\begin{figure}
{\par\centering \resizebox*{4.5cm}{5.5cm}{\includegraphics{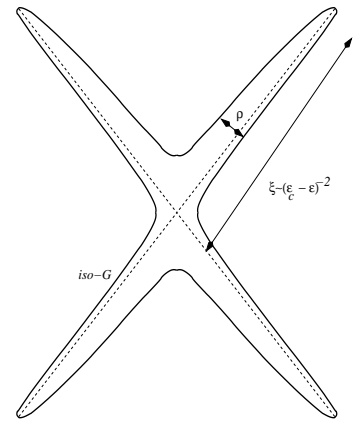}} \par}
\caption{Form of an iso-autocorrelation curve in the approach to the localization
transition. }
\end{figure}
This prediction can be interpreted as corresponding to the formation of clusters
of microcracks having  aspect ratios  \( \xi /\rho  \) that diverge 
as the cracks organize into  long thin structures  along which the sample will 
ultimately fail to form the experimentally observed shear bands.

\subsection{Experimental measurement of \protect\( \xi \protect \)}

It would certainly be desirable to have direct experimental verification of
whether the crack bands have aspect ratio that diverge as \( 1/(\varepsilon _{c}-\varepsilon )^{2} \).
Unfortunately, there are many practical problems that have prevented the direct
measurement of the autocorrelation function of cracks in materials like rocks.
We comment here on three types of measurements that either have or could be
used to quantify the autocorrelation.

First, following ideas used by Davy and Bonnet \cite{DB98} in interpreting
their sandbox shear experiments, one  
can  measure the local deformation of a large sample by covering
the surface with pixels and montoring the shear strain of
each pixel. The total shear strain of the system is then approximated by taking
the average over the surface pixels. If the  system deformation
is plotted as a function of the pixel size, it is expected that
when the pixels are smaller than the emergent band structures, the system deformation
will decrease as a powerlaw of increasing pixel size as  was observed by Bonnet and Davy. 
 However, at a particular pixel  
size there is a cross over to a constant system deformation as pixel size increases.
The pixel size at the cross-over point is at least an indirect measurement of
the correlation length \( \xi  \) above which a volume-averaged description
of the system holds with properties independent of the pixel size.

Second, a direct measurement of the autocorrelation between cracks can in principal
be obtained via acoustic-emissions monitoring \cite{Loc95}. However, the present
resolution of this method (millimeters in centimeter-scale specimens) and the
difficulty in determing the mode of the individual crack events prevents having
a satisfactory sampling for statistical analysis. It seems that improvements
on these present limitations are possible.

Last, by analogy with the probing of spin populations by electromagnetic waves
to study the ferro/paramagnetic transitions, it should be possible to send plane
sound waves through a system and measure the scattering cross-section as the
waves scatter from the structure of the evolving microcrack population. We have
not yet obtained the rigorous connection between such a measured cross-section
and the Fourier-transform of our autocorrelation function; however, such a relation
almost certainly exists. No experimental attempts to measure the correlation
function of cracking systems in this manner has been attempted to our knowledge.

\section{Conclusion}
We now summarize the principal results that have emerged in our study.  
First, we have demonstrated that at a well-defined strain point $\omega=\omega_c$, 
thin bands of coherently oriented cracks can be added to the system  at no 
energetic cost.  Such localized structures break the symmetry that held 
when $\omega < \omega_c$ and   correspond to a phase transition that 
we named the ``localization transition''.  
It was demonstrated that the free energy $F$ and entropy of the system remain 
continuous and finite at  the localization transition which   
 justifies calling it a critical-point phenomena.  Such continuity also 
demonstrates that the stress/strain behaviour of the rock is entirely analytic 
up to and including localization.   
The only divergence at localization is in the second derivatives of $F$ with respect 
to the external field $J$.   
 The consequence is that the correlation 
length (aspect ratio) of the emergent crack clusters  diverges as $(\omega_c - \omega)^{-2}$.  
Presumably, if the ``mean-phase'' approximation had not been invoked and if 
order-parameter contributions proportional to  $\varphi^3$ and higher had been retained in the 
Hamiltonian through a 
renormalization scheme,  
 then a non-trivial exponent on this scaling law might emerge. 

The mechanical behavior of the system at localization exhibits many  qualities 
observed in  actual experiments on rocks.  First, the stress components 
at localization are reduced relative to their values if the rock had remained intact. 
The total dilatation $\Delta_c$  remains negative at localization, even though the radial strain is
 positive.
With  radial confining stress  kept constant, 
the tangent moduli $d\sigma /d\varepsilon$ are, most normally, 
 negative at localization indicating 
that the load curve has already gone through a smooth quadratic peak stress prior to localization. 
 Nonetheless, for rocks with a sufficiently low bulk modulus and 
at sufficiently low confining pressures, the localization can occur in the hardening regime, 
presumably followed by a sharp peak stress corresponding to the unstable coalescence 
of  cracks as the sample fails along a shear band. 
These results are consistent with what experimentalists observe.   

Using the exact differential equation that controls the temperature in the theory, 
it has been demonstrated that the temperature  is  becoming even  more 
negative at localization which means that the temperature is always finite at localization. 
Unfortunately, the exact  value $T_c$ of the temperature at localization is 
difficult to obtain because it is a result of integrating the differential equation 
from the initial conditions.  Although this could be done  numerically, 
we have instead used an approximate value of $T_c$ based on a non-interacting 
crack model.  

By far the most important signature of the localization  transition is the 
divergence of the aspect ratio of the crack clusters.  As reported, no definitive 
experimental work has yet been performed to test this  prediction and we 
hope that experimentalists  take this  as a  challenge.

\bibliographystyle{apsrev}

\end{document}